\documentclass{ws-ijmpa}
\usepackage[super,compress]{cite}
\usepackage{graphicx}
\usepackage{physics}
\usepackage{enumitem}
\usepackage{nicefrac}
\usepackage{bbold}
\usepackage{mathtools}
\usepackage[hidelinks]{hyperref}

\DeclareMathOperator{\0}{\mathbb{0}}
\DeclareMathOperator{\1}{\mathbb{1}}
\DeclareMathOperator{\diag}{diag}

\newcommand{\Var}{\operatorname{Var}} 
\newcommand{\obs}{\mathcal{O}} 
\newcommand{\phaseD}{{\updownarrow}}
\newcommand{\phaseS}{{\uparrow\downarrow}}
\newcommand{\phaseO}{{\uparrow\uparrow}}

\def\lr#1{\left(#1\right)}

\makeatletter
\DeclareRobustCommand{\cev}[1]{%
  {\mathpalette\do@cev{#1}}%
}
\newcommand{\do@cev}[2]{%
  \vbox{\offinterlineskip
    \sbox\z@{$\m@th#1 x$}%
    \ialign{##\cr
      \hidewidth\reflectbox{$\m@th#1\vec{}\mkern4mu$}\hidewidth\cr
      \noalign{\kern-\ht\z@}
      $\m@th#1#2$\cr
    }%
  }%
}
\makeatother

\begin{document}
\markboth{
D. Prekrat, 
D. Rankovi\'{c},  
N. K. Todorovi\'{c}-Vasovi\'{c}, 
S. Kov\'{a}\v{c}ik, 
and J. Tekel}{Phase transitions in a $\Phi^4$ matrix model on a curved noncommutative space}

%
\catchline{}{}{}{}{}
%

\title{Phase transitions in a $\Phi^4$ matrix model on a curved noncommutative space}

\author{Dragan Prekrat\footnote{speaker at COST Action CA18108: Workshop on theoretical aspects of quantum gravity, 1-3 September 2022, Belgrade}}

\address{Faculty of Pharmacy, University of Belgrade, Vojvode Stepe 450\\
Belgrade, 11221, Serbia\\
dprekrat@ipb.ac.rs}

\author{Dragana Rankovi\'{c}}

\address{Faculty of Pharmacy, University of Belgrade, Vojvode Stepe 450\\
Belgrade, 11221, Serbia\\
dragana.rankovic@pharmacy.bg.ac.rs}

\author{Neli Kristina Todorovi\'{c}-Vasovi\'{c}}

\address{Faculty of Pharmacy, University of Belgrade, Vojvode Stepe 450\\
Belgrade, 11221, Serbia\\
kristina.todorovic@pharmacy.bg.ac.rs}

\author{Samuel Kov\'{a}\v{c}ik}

\address{Department of Theoretical Physics, Faculty of Mathematics, Physics and Informatics, Comenius University in Bratislava, Mlynsk\'a dolina, 842 48 \\
Bratislava, Slovakia\\
Department of Theoretical Physics and Astrophysics, Faculty of Science, Masaryk University, Kotl\'a\u{r}sk\'a 267/2, Veve\u{r}\'{i} \\
Brno, Czech Republic\\
samuel.kovacik@fmph.uniba.sk}

\author{Juraj Tekel}

\address{Department of Theoretical Physics, Faculty of Mathematics, Physics and Informatics, Comenius University in Bratislava, Mlynsk\'a dolina, 842 48 \\
Bratislava, Slovakia\\
juraj.tekel@fmph.uniba.sk}

\maketitle

\begin{history}
\received{Day Month Year}
\revised{Day Month Year}
\end{history}

\begin{abstract}
In this contribution, we summarize our recent studies of the phase structure of the Grosse-Wulkenhaar model and its connection to renormalizability. Its action contains a special term that couples the field to the curvature of the noncommutative background space. We first analyze the numerically obtained phase diagram of the model and its three phases: the ordered, the disordered, and the noncommutative stripe phase. Afterward, we discuss the analytical derivation of the effective action and the ordered-to-stripe transition line, and how the obtained expression successfully explains the curvature-induced shift of the triple point compared to the model without curvature. This shift also causes the removal of the stripe phase and makes the model renormalizable.

\end{abstract}

\ccode{PACS numbers: 02.10.Yn, 03.65.Vf, 05.10.Ln, 04.62.+v, 04.60.-m}



\section{Introduction}

One plausible way of bringing the domains of gravitation and quantum theory, which are reluctant to cooperate, closer together is to modify \cite{Hossenfelder:2012jw} the short-distance structure of spacetime, for example by introducing coordinate noncommutativity (NC). This, unexpectedly, complicates the renormalization of NC models due to UV/IR mixing \cite{Minwalla:1999px} which intertwines physics at large and small scales and interferes with the scale separation. 

Grosse-Wulkenhaar (GW) model \cite{Grosse:2003nw} is a prototype for a renormalizable NC model without UV/IR mixing problems. It features an additional harmonic-potential $\Omega$-term that introduces symmetry between the scales through the Langman-Szabo duality \cite{Langmann_2002}. This term also has a nice geometric interpretation --- a coupling with the curvature of the background NC space of the truncated Heisenberg algebra \cite{Buric:2009ss} $\mathfrak{h}^\text{tr}$.

Recently, the renormalizability of the GW model has been connected to the removal of the NC striped phase from its phase diagram. This was done by analyzing its matrix field formulation, both analytically \cite{Prekrat:2022sir} and numerically \cite{Prekrat:2021uos,Prekrat:2020ptq}. This contribution summarizes our results obtained on these fronts in the previous years.


\section{GW model}

Let us first introduce our main actor and set up the NC stage. The two-dimensional GW model is given by \cite{Grosse:2003nw}
\begin{multline}
S_\text{\tiny GW} = \int dx^2
\Bigg(
\frac{1}{2} \partial^\mu\phi \star \partial_\mu\phi
+ \frac{\Omega^2}{2}((\theta^{-1})_{\mu\rho}x^\rho\phi) \star ((\theta^{-1})^{\mu\sigma}x_\sigma\phi) +
\\
+ \frac{m^2}{2}\phi\star\phi
+ \frac{\lambda}{4!}\phi\star\phi\star\phi\star\phi
\Bigg),
\label{GW model}
\end{multline}
and it lives on the Moyal plane equipped with a $\star$-product and NC coordinates:
\begin{equation}
    f \star g= f\,e^{\nicefrac{i}{2}\,\cev{\partial}_\mu\theta^{\mu\nu}\vec{\partial}_\nu}\,g
\quad\Rightarrow\quad
    \comm{x^\mu}{x^\nu}_\star=i\theta\epsilon^{\mu\nu}.
\end{equation}
Thanks to the $\Omega$-term, this model is superrenormalizable \cite{Wulkenhaar:habilitation2004}; without it, we are left with the nonrenormalizable $\lambda\phi^4_\star$ model.

Using the Weyl transform and a slight adjustment of coefficients, \eqref{GW model} can be turned into a matrix model 
\begin{equation}
S = N\tr\lr{
\Phi\mathcal{K}\Phi
- g_r R\Phi^2
- g_2\Phi^2 + g_4\Phi^4
},
\label{GW matrix model}
\end{equation}
where the field $\phi$ becomes a $N \times N$ Hermitian matrix $\Phi$. In addition to regularizing the model, the finite matrices are also simulation-friendly and allow a glimpse into nonperturbative results. The model's curved background space $\mathfrak{h}^\text{tr}$ is now spanned by rescaled NC coordinates represented by matrices
\\
\begin{equation}
X = \frac{1}{\sqrt{2N}}
\begin{pmatrix} 
 & \scriptstyle+\sqrt{1} \\ 
 \scriptstyle+\sqrt{1} &  & \scriptstyle+\sqrt{2} \\ 
 & \scriptstyle+\sqrt{2} & &  \hspace{-25pt}\begin{rotate}{-5}{$\ddots$}\end{rotate} \\ 
 &  & \hspace{-25pt}\begin{rotate}{-5}{$\ddots$}\end{rotate} &  & \hspace{-16pt}\scriptstyle+\sqrt{N-1} \\ 
 &  & & \hspace{-15pt}\scriptstyle+\sqrt{N-1} &
\end{pmatrix},
\quad
Y = \frac{i}{\sqrt{2N}}
\begin{pmatrix} 
 & \scriptstyle-\sqrt{1} \\ 
 \scriptstyle+\sqrt{1} &  & \scriptstyle-\sqrt{2} \\ 
 & \scriptstyle+\sqrt{2} & & \hspace{-25pt}\begin{rotate}{-5}{$\ddots$}\end{rotate} \\ 
 &   & \hspace{-25pt}\begin{rotate}{-5}{$\ddots$}\end{rotate} &  & \hspace{-16pt} \scriptstyle-\sqrt{N-1} \\ 
 &  &    & \hspace{-15pt}\scriptstyle+\sqrt{N-1} &
\end{pmatrix},
\end{equation}
\\
\noindent
and its rescaled curvature $R$ hides the energy levels of the $\Omega$-term harmonic oscillator
\begin{equation}
R=\frac{15}{2N}-8\lr{X^{2}+Y^{2}} \stackrel{\scriptscriptstyle N\to\infty}{\approx} -\frac{16}{N} \diag\lr{1,2,\ldots N}.
\end{equation}
In the infinite-size limit, we recover the commutation relations of the Moyal plane.
Finally, the differential calculus is defined \cite{Buric:2009ss} in such a way that coordinates also act as momenta, leading to the kinetic operator given by a double commutator
\begin{equation}\label{kinetic}
\mathcal{K}\Phi = \comm{X}{\comm{X}{\Phi}} + \comm{Y}{\comm{Y}{\Phi}} \, .
\end{equation}

We have absorbed the NC scale $\theta$ into definitions of matrices and couplings and set it to $\sqrt{\theta}$ = 1, which allows us to work with dimensionless quantities. It will also be useful to keep an eye on the unscaled\footnote{containing the factor $N$ from the action \eqref{GW matrix model}.} couplings
\begin{equation}
    G_2 = Ng_2 \, ,
    \qquad
    G_4 = Ng_4 \, ,
\end{equation}
as they appear in the analytical results for renormalization.

Having defined our matrix action, we can now ask questions about the observables $\obs$ of interest and have them answered with the help of well-defined matrix path integrals:
\begin{equation}
    \expval{\obs} = \frac{\displaystyle\int [d\Phi] \, \obs \, e^{-S}}{\displaystyle\int [d\Phi] \, e^{-S}} \, ,
    \qquad\quad
    \Var\obs=\expval{\obs^{\smash{2}}}-{\expval{\obs}}{^2} \, .
\end{equation}

\section{Phase diagram}

Since we are interested in the phase structure of the GW model, it is instructive to look at the possible variants for its classical vacua. The saddle point method gives us the classical EOM
\begin{equation}
2\mathcal{K}\Phi-g_r \lr{R \Phi + \Phi R}
+ \Phi\lr{-2g_2+4g_4\Phi^2} = 0 \, ,
\end{equation}
whose solutions corresponding to the kinetic ($\mathcal{K}$), curvature ($R$), and pure-potential\footnote{We are interested in the $g_2>0$ regime with the spontaneous symmetry breaking.} ($g_2,g_4$) parts are given, respectively, by
\begin{equation}\label{vacuum solutions}
\;\;
\Phi_\phaseO = \frac{\tr\Phi}{N}\1 \, ,
\qquad\quad
\Phi_\phaseD = \0 \, ,
\qquad\quad
\Phi^2_\phaseS = \frac{g_2}{2g_4} \1 \, .
\end{equation}
The first two are the ordered ($\phaseO$) and the disordered ($\phaseD$) vacuums found also in commutative models; the third is NC-specific, since nontrivial square roots of the identity matrix can occur in the case of spontaneous symmetry breaking, giving rise to the so-called stripe phase ($\phaseS$). The stripe phase breaks the translational symmetry since it has both positive and negative eigenvalues and thus varies \cite{Castorina:2007za,Mejia-Diaz:2014lza,Ambjorn:2002nj} throughout space.

\begin{figure}[t]
\centering
\includegraphics[width=6.2cm]{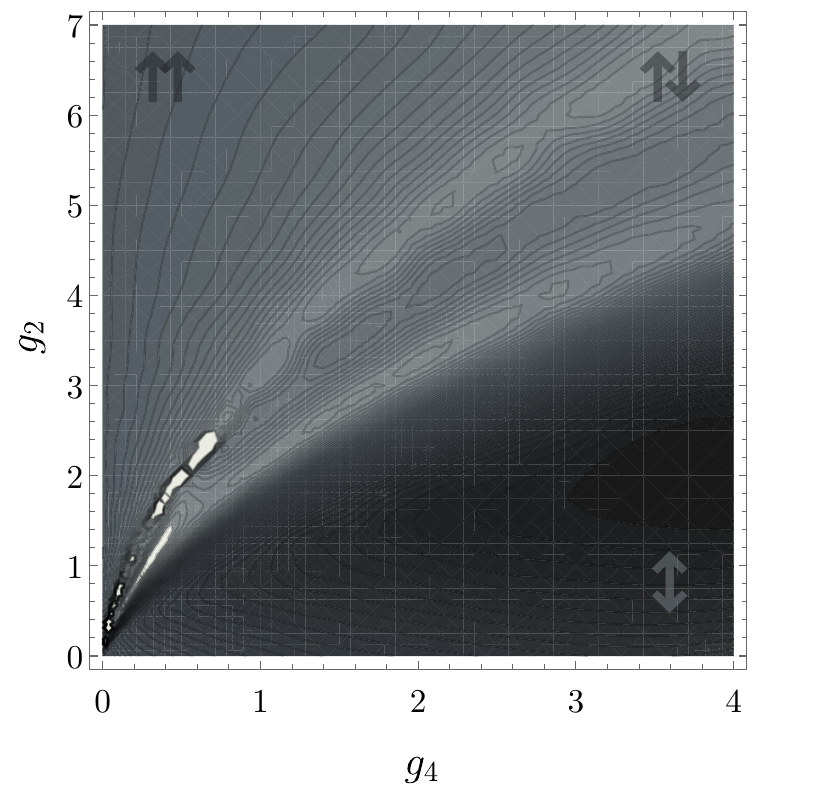}
\includegraphics[width=6.2cm]{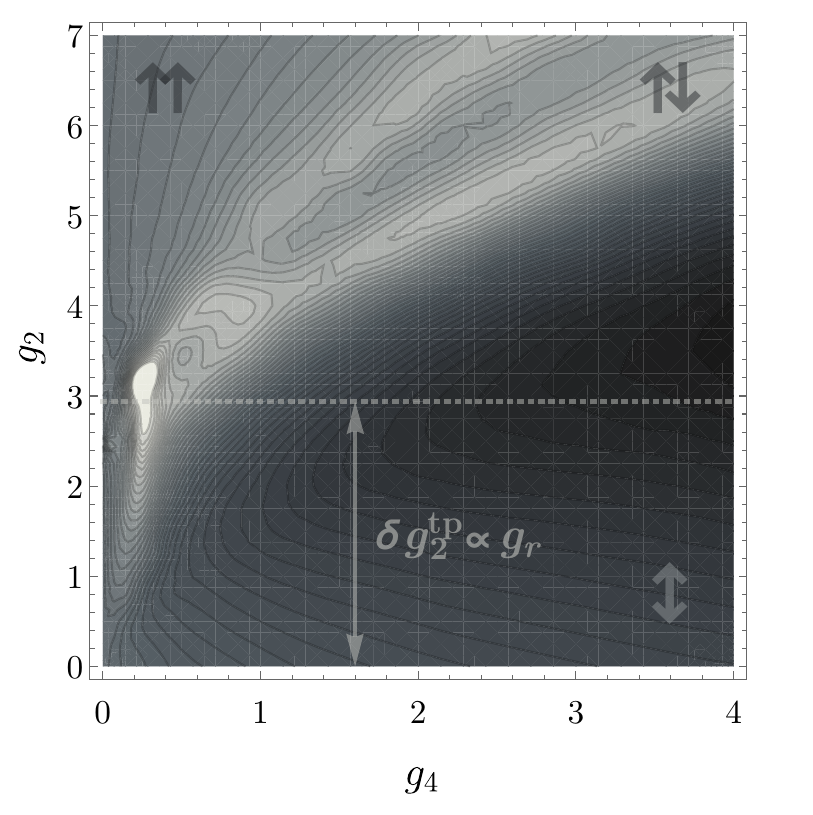}
\caption{Phase diagrams for the model with $g_r = 0$ (left) and $g_r = 0.2$ (right). The two diagonal bright stripes on each of the plots represent the transition lines. The $\phaseD$-phase occupies the bottom right, the $\phaseO$-phase the top left, while the NC $\phaseS$-phase is sandwiched in-between. The diagrams are constructed from contour plots of the specific heat for $N=24$ and represent the updated and modified version of Fig. 1 from Ref. \citen{Prekrat:2021uos}.\label{phase diagrams}} 
\end{figure}

As shown in Fig. \ref{phase diagrams}, such a phase structure is precisely what we found \cite{Prekrat:2022sir,Prekrat:2021uos,Prekrat:2020ptq} using numerical Hamiltonian Monte Carlo simulations \cite{Ydri:2015zba}, aided occasionally by an eigenvalue-flipping algorithm \cite{Kovacik:2022kfh}. These phase diagrams are constructed for both the model with (GW) and without ($\lambda\phi^4_\star$) curvature effects by scanning the $(g_2,g_4)$ parameter space for peaks and edges in profiles of specific heat $C = \Var S/N$, which are some possible signatures of a phase transition. Darker regions in the diagrams have lower values and lighter regions have higher values of $C$. In each of the diagrams, two bright stripes rising from the common triple point from left to right thus represent the phase-transition lines separating the three phases indicated by the arrows in the corners of the diagrams. These lines follow \cite{Prekrat:2022sir} a (shifted) deformed square-root law, the exact version of which is found in the pure-potential model \cite{Filev:2014jxa,Tekel:2017nzf}. As can be seen, the $\phaseD$-phase occupies the lower right corner due to the stronger $g_4$-interaction, the $\phaseO$-phase is in the upper left corner where the mass parameter $g_2$ is greater, and the $\phaseS$-phase is sandwiched in-between. This diagram structure is also found in matrix models on other NC spaces such as the fuzzy sphere \cite{Kovacik:2018thy,Ydri:2014uaa,Medina:2007nv}, the fuzzy disc \cite{Rea:2015wta}, and the fuzzy complex projective space \cite{Tekel:2014bta,Saemann:2010bw}.

The extensive analysis of various other observables, different matrix sizes and parameter scalings, as well as the details on the numerical method and algorithms, can be found in Ref. \citen{Prekrat:2021uos} and Ref. \citen{Prekrat:2020ptq}.


\section{Renormalization}

As advertized in the introduction, the extension of the $\phaseS$-phase in the phase diagram of the NC model seems to be related to the renormalizability of the model. We will now clarify the details of this connection.

There are several important ingredients that tie these properties together. The first of these is shown in Fig. 1. Namely, if we compare the two phase diagrams with $g_r = 0$ and $g_r \neq 0$, we see that there is an overall shift of the latter towards the larger values of $g_2$ that is proportional to the curvature coupling $g_r$. More precisely, we have found \cite{Prekrat:2022sir} that the triple point of the $\lambda\phi^4_\star$ model is glued to the origin in the large-$N$ limit while the triple point of the GW model is separated from the origin by
\begin{equation}
    \delta g_2^\text{tp} \propto g_r \, .
\end{equation}

Since we often calculate the renormalization of the model in unscaled parameters, we can translate this picture from $(g_2,g_4)$ to $(G_2, G_4)$. This is our second ingredient. Since this change stretches the diagram axis, it affects the large-$N$ limits of the diagrams. It turns out that after this stretch the triple point of $\lambda\phi^4_\star$ still remains at the origin while the triple point of the GW model now shifts to infinity and out of the diagram as
\begin{equation}
    \delta G_2^{\,\text{tp}} = N\,\delta g_2^\text{tp} \propto Ng_r \to \infty,
    \qquad
    N \to \infty \, .
\end{equation}
Thus, with respect to the unscaled parameters, the nonrenormalizable $\lambda\phi^4_\star$ model is left with the $\phaseS$-phase, while in the renormalizable GW model, the $\phaseS$-phase is completely removed. 

Furthermore, if we consider the mass renormalization \cite{Vinas:2014exa} in the GW model
\begin{equation}
    \delta m^2_\text{ren} = \frac{\lambda}{12\pi(1+\Omega^2)}\log\frac{\Lambda^2\theta}{\Omega} \, ,
\end{equation}
using the cutoff \cite{Grosse:2003nw} $\Lambda^2 \propto N$, and translating this into the language of the matrix model parameters, we obtain
\begin{equation}
    \delta G_2^\text{\,ren} \sim -\log N,
\end{equation}
hence the bare $G_2$ must increase by $\abs{\delta G_2^{\,\text{ren}}}$ to compensate for the quantum effects, shifting accordingly its position in the phase diagram.
Since $\abs{\delta G_2^{\,\text{ren}}} < \delta G_2^{\,\text{tp}}$, this means that the bare mass of the model remains in the $\phaseD$-phase and consequently out of the $\phaseS$-phase in the large-$N$ limit. 

We know \cite{Grosse:2003nw, Wulkenhaar:habilitation2004} that the GW model can be used to construct the renormalizable version of the $\lambda\phi^4_\star$ by turning off the curvature coupling sufficiently slowly as we increase $N$. Indeed, since it does not renormalize, $\Omega$ can serve as a label for a series of superrenormalizable models $\text{GW}_\Omega$ whose limit we will denote by $\lambda\phi^4_\text{\tiny GW}$:
\begin{equation}
    \text{GW}_\Omega \to \lambda\phi^4_\text{\tiny GW},
    \qquad
    \Omega \sim \frac{1}{\log N} \to 0 \, .
\end{equation}
The comparison between the $\lambda\phi^4_\star$ and $\lambda\phi^4_\text{\tiny GW}$ phase diagrams will be our final ingredient in the argument about the connection between the $\phaseS$-phase and (non)renormalizability. When we change to matrix-model parameters, $\Omega \sim 1/\log N$ becomes \cite{Prekrat:2021uos} $g_r \sim 1/\log^2 N$, and accordingly 
\begin{equation}\label{triple point shift}
    \delta G_2^{\,\text{tp}} \propto Ng_r \sim \frac{N}{\log^2 N} \to \infty,
    \qquad
    N \to \infty \, .
\end{equation}
Therefore, although formally both $\lambda\phi^4_\star$ and $\lambda\phi^4_\text{\tiny GW}$ lack the curvature term, the former/nonrenormalizable has the $\phaseS$-phase tethered to the phase-diagram origin while the latter/renormalizable is $\phaseS$-phase-free.

Looking once more at the mass renormalization, we see that the leading divergence
\begin{equation}
    g_r \sim \frac{1}{\log^2 N}
    \quad\Rightarrow\quad
    \abs{\delta G_2^{\,\text{ren}}} \sim \log N \, ,
\end{equation}
still grows slower than the triple-point shift \eqref{triple point shift} and that the bare mass still remains outside the $\phaseS$-phase.


\section{Analytical results}

Since the conclusions of the previous section hinge on the numerically obtained phase-diagram shift, it would be desirable to derive this shift by analytical means. This is partially achieved in Ref. \citen{Prekrat:2022sir}. 

As can be seen from the classical vacuum solutions \eqref{vacuum solutions}, depending on the shape and depth of the potential well, their eigenvalues can be all around zero, all around one minimum of the potential well, or simultaneously around both the positive and negative minimum. This means that the eigenvalue distributions in the different phases can be classified according to their symmetry and the connectedness of their support: the $\phaseD$-phase has a symmetric one-cut distribution, the $\phaseO$-phase has an asymmetric one-cut distribution, while the $\phaseS$-phase is described by a two-cut distribution. 

To obtain these distributions and the conditions for their existence, and to extract from them the positions of the phase-transition lines, we must first integrate out the angular degrees of freedom in the partition function $Z$ coming from the unitary matrices $U$ in the field decomposition $\Phi=U\Lambda U^\dag$, and obtain the effective action that depends only on the eigenvalues $\Lambda = \diag \lambda_i$. Because of the double commutators in the kinetic operator $\mathcal{K}=[X,[X,\cdot\,]] + [Y,[Y,\cdot\,]]$, unlike in the potential term, unitary matrices cannot be canceled out under the trace, leading to the complicated angular integral. Since numerical simulations \cite{Prekrat:2020ptq,Prekrat:2021uos} imply that the curvature gives the main contribution to the shift, while the kinetic term merely turns on the additional $\phaseO$-phase, we will drop the kinetic term and focus on the slightly less complicated curvature contribution
\begin{equation}\label{partition function}
    Z = \int[d\Phi] \, e^{-S} = 
    \int [d\Lambda] \, \Delta^2(\Lambda) \, 
    e^{-N\tr\left(- g_2\Lambda^2 + g_4\Lambda^4\right)}
\int [dU]\,e^{\, g_r N\tr\lr{URU^\dagger \Lambda^2}} \, ,
\end{equation}
where $\Delta(\Lambda)$ is a Vandermonde determinant that depends only on the argument's eigenvalues
\begin{equation}
    \Delta(\Lambda) = \prod_{1 \leq i < j \leq N} (\lambda_j - \lambda_i) \, ,
\end{equation}
and appears from the Jacobian of the change of variables in the integral.
If we use the method described in Ref. \citen{Kanomata:2022pdo} and Ref. \citen{Kanomata:2023mni}, where the exact analytical treatment of the correlation functions for the $\Phi^3$ and hybrid $\Phi^3$/$\Phi^4$ NC models is given, we can perform \cite{Prekrat:2022sir} the $[dU]$-integration to obtain
\begin{equation}
Z= \int [d\Lambda] \, \frac{\Delta^2(\Lambda)}{\Delta(\Lambda^2)} \,
e^{-N\tr(-g_2\Lambda^2 - g_r R\Lambda^2 + g_4 \Lambda^4)}\, ,   
\end{equation}
however, since $\Delta(\Lambda^2)$ does not have a definite sign for all $\Lambda$, we cannot perform the final step and absorb it into the effective action in the exponent. 

It turns out that \eqref{partition function} can be instead tackled perturbatively with the help of the Harish-Chandra-Itzykson-Zuber integral \cite{Zinn-Justin:2002rai}
\begin{equation}\label{HCIZ-integral}
     I = \int\limits_{{\rm U}(N)} \!\! [dU]\, e^{t\tr (AUBU^\dag)} 
    = \frac{c_N \, t^{-N(N-1)/2}}{\Delta(A)\Delta(B)}
    \begin{vmatrix} 
    e^{ta_1b_1} & e^{ta_1b_2} & \cdots & e^{ta_1b_N} 
    \\
    e^{ta_2b_1} & e^{ta_2b_2} & \cdots & e^{ta_2b_N} 
    \\
    \vdots & \vdots & & \vdots 
    \\
    e^{ta_Nb_1} & e^{ta_Nb_2} & \cdots & e^{ta_Nb_N}
\end{vmatrix},
\end{equation}
where $c_N$ is a constant that depends on the matrix size. The trick is to expand the determinant on the right-hand side of \eqref{HCIZ-integral} in powers of $t$ and then re-exponentiate it into the effective action $S(\Lambda)$, which contains only the field eigenvalues. This was done in Ref. \citen{Prekrat:2022sir} up to $O(g_r^4)$, leading to the multitrace action
\begin{multline}\label{effective action}
     S(\Lambda) = 
    - \lr{g_2 - 8g_r}N\tr\Lambda^2 
    + \lr{g_4 - \frac{32}{3}g_r^2}N\tr\Lambda^4 
    + \frac{1024}{45}g_r^4N\tr\Lambda^8 
    +
    \\
    + \frac{32}{3}g_r^2\tr^{\mathrlap{2}}\, \Lambda^2
    + \frac{1024}{15}g_r^4\tr^{\mathrlap{2}}\,\Lambda^4
    - \frac{4096}{45}g_r^4\tr\Lambda^6\tr\Lambda^2
    - \log\Delta^2(\Lambda) \, .
\end{multline}
As we can see from the first term in the action, the leading effect of the curvature is to shift the mass parameter by $8g_r$, which is proportional to the curvature coupling $g_r$, as expected, but only by half the amount found for the triple-point shift in numerical simulations \cite{Prekrat:2021uos}. This is a good place to notice \cite{Prekrat:2020ptq} that if we replace $R$ in \eqref{GW matrix model} by its eigenvalue with the largest absolute value, we can expect the triple-point shift to be at most $16g_r$ in the large-$N$ limit, so the simulations imply that this bound is saturated.

\begin{figure}[t]
\centering
\includegraphics[width=6.0cm]{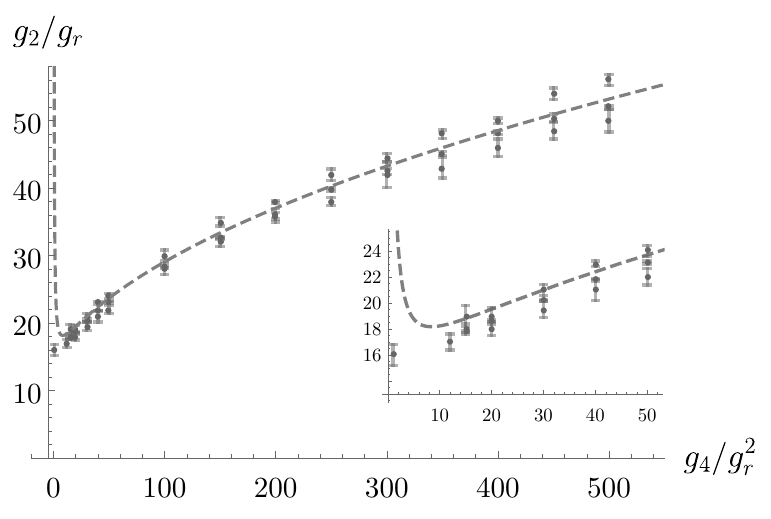}
\includegraphics[width=6.0cm]{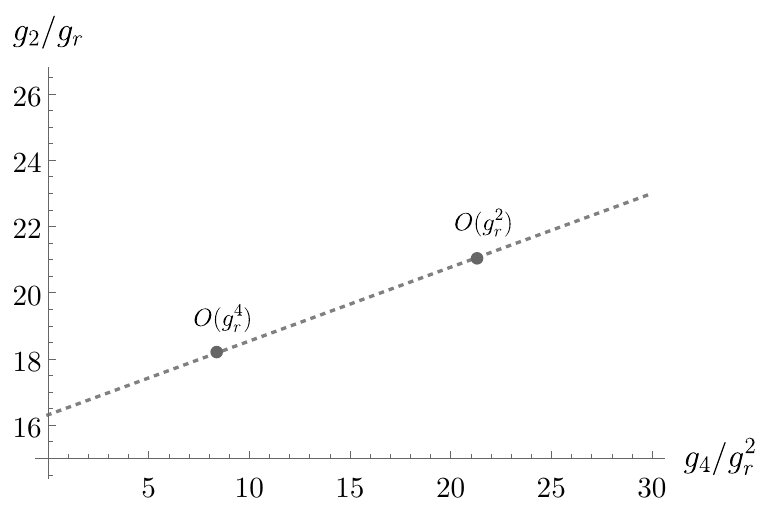}
\caption{(left) Comparison of the dashed analytical line \eqref{analytical line} and numerical points with errorbars for the $\phaseO\to\phaseS$ transition at $N=24$ and $g_r=0.1$. The plot is the modified version of Fig. 5 from Ref. \citen{Prekrat:2021uos}. (right) Turning points of the transition line at $O(g_r^2)$ and $O(g_r^4)$ where the dotted-line ending near $(g_2,g_4)=(16g_r,0)$ represents a naive extrapolation to higher orders. \label{transition line}} 
\end{figure}

Similarly to other matrix models \cite{Filev:2014jxa,Tekel:2017nzf,Tekel:2014bta,Tekel:2015uza,Subjakova:2020shi,Ydri:2014zya}, by using the saddle point method and taking the continuous limit $N \to \infty$, we arrive at the equation for the eigenvalue distribution $\rho(\lambda)$,
\begin{multline}\label{distribution equation}
    -\lr{g_2 - 8g_r - \frac{64}{3}g_r^2m_2 + \frac{4096}{45}g_r^4 m_6}\lambda 
    +2\lr{g_4 - \frac{32}{3}g_r^2 + \frac{2048}{15}g_r^4 m_4}\lambda^3
    -
    \\
    -\frac{4096}{15}g_r^4 m_2\lambda^5
    +\frac{4096}{45}g_r^4 \lambda^7
    =
    \underset{\mathclap{\text{support}}}{\int} d\lambda' \, \frac{\rho(\lambda')}{\lambda-\lambda'} \, ,
\end{multline}
which also depends on the moments of the distribution
\begin{equation}
    m_k = \underset{\mathclap{\text{support}}}{\int}  d\lambda' \, \rho(\lambda')\, \lambda'^{\,k} .
\end{equation}
The equation \eqref{distribution equation} can be solved \cite{Prekrat:2022sir} order-by-order in $g_r$ with appropriate ansatz, and from the one-cut solution existence condition we have found that the $\phaseO\to\phaseS$ transition line satisfies
\begin{equation}\label{analytical line}
    g_2 = 2\sqrt{g_4} + 8g_r + \frac{32}{3}\frac{g_r^2}{\sqrt{g_4}} + \frac{256}{15}\frac{g_r^4}{g_4\sqrt{g_4}} \, ,
\end{equation}
up to $O(g_r^4)$.
As the left diagram in Fig. \ref{transition line} shows, \eqref{analytical line} cannot be trusted near $g_4 = 0$, where it deviates starkly from the numerical results, which means that the would-be-triple-point shift is not fully described by $\delta g_2^\text{tp} = 8g_r$. However, looking at the right plot and the positions of the transition-line turning points for $O(g_r^2)$ and $O(g_r^4)$, where the approximation starts to fail, we notice that they move towards the desired point $(g_2,g_4)=(16g_r,0)$, i.e. the shift $\delta g_2^\text{tp} = 16g_r$, when higher terms are included in the effective action.


\section{Conclusions and outlook}

In this contribution, we have compiled the results of the recent investigation of the phase structure of the GW model and its connection to renormalizability. The phase diagram is found to be divided into three phases, which are also present in other NC models: the ordered, the disordered, and the NC stripe phase. We have also sketched the derivation of the ordered-to-stripe transition line in the strong interaction regime without the kinetic term and shown that the expression successfully explains the curvature-induced shift of the triple point. By expressing the results in unscaled parameters, we have shown that this shift causes the removal of the stripe phase in the renormalizable GW model and also in the renormalizable $\lambda\phi^4_\text{\tiny GW}$ model obtained by slowly turning off the curvature coupling, but not in the original nonrenormalizable $\lambda\phi^4_\star$ without the curvature.

It would be interesting to investigate whether this connection between phase structure and renormalizability also holds in other NC models. For example, one could try to simulate the related model \cite{Franchino-Vinas:2021bcl,Franchino-Vinas:2019nqy} on the Snyder-de Sitter space that naturally develops the GW harmonic oscillator term, either in the matrix form (if possible) or on a lattice \cite{Mejia-Diaz:2014lza,Szabo:2001kg}. Another promising option would be the nonrenormalizable $U(1)$ gauge model \cite{Buric:2016lly} on $\mathfrak{h}^\text{tr}$. This model has two possible classical vacua: the trivial zero-vacuum and the stripe vacuum proportional to the NC coordinates. Numerical simulations are needed to determine which of these is actually realized in the model. It would also be nice to study the correlation functions of the GW model in a similar way as it is done \cite{Hatakeyama:2018qjr} on the fuzzy sphere. 

On the analytical side, we could try adding $O(g_r^6)$ terms to the effective action \eqref{effective action}, as this would be a straightforward, if somewhat more involved, continuation of the existing approach, and see if the turning point of the transition line follows the trend from Fig. \ref{transition line}. This would give us the first nontrivial extrapolation of the starting point of the transition line. Finally, we would also like to find a way to extend the analytical treatment of the curvature term to the kinetic term of the GW model.


\section*{Acknowledgments}

This research was funded by:
\begin{itemize}[leftmargin=25pt]
    \item the Ministry of Education, Science and Technological Development, Republic of Serbia through:
    \begin{itemize}
    \item
    Grant Agreement with University of Belgrade -- Faculty of Pharmacy No. 451-03-68/2022-14/200161,
    \item Grant Agreement with University of Belgrade -- Faculty of Physics No. 451-03-68/2022-14/200162,
    \end{itemize}
    \item the Ministry of Science, Technological Development and Innovation, Republic of Serbia through Grant Agreement with University of Belgrade -- Faculty of Pharmacy No. 451-03-47/2023-01/200161,
    \item VEGA 1/0703/20 grant \emph{Quantum structure of spacetime}. 
    \item VEGA 1/0025/23 grant
\end{itemize}


\bibliographystyle{ws-ijmpa}
\bibliography{bibliography.bib}

\end{document}